\documentclass[
    ,final            
  ]
  {aipproc}

\layoutstyle{6x9}
\usepackage{atlasphysics}

\begin{document}

\title{W and Z Boson Production Cross Section Measurements in ATLAS}

\classification{14.70.Fm, 14.70.Hp}
\keywords      {W/Z Production, W Cross Section, Z Cross Section}

\author{Verena I. Martinez Outschoorn$^*$ on behalf of the ATLAS Collaboration}{
  address={$^*$Harvard University, 17 Oxford St. Cambridge MA 02138, vimartin@physics.harvard.edu}
}

\begin{abstract}
The electroweak boson inclusive production cross sections and their ratios are described in 
the electron and muon decay channels. The observation of $W$ and $Z$ bosons in the tau 
decay channel is also presented. The data from $pp$ collisions at 7 TeV were collected during 
$2010$ using the ATLAS detector at the Large Hadron Collider and the results are compared 
to predictions at NNLO in perturbative QCD.
\end{abstract}

\maketitle


\section{Introduction}

Measurements of the production of $W$ and $Z$ bosons probe perturbative 
QCD predictions, testing the validity of theoretical NNLO calculations and providing
information on the parton distribution functions (PDFs). The results presented here~\cite{wzconf} 
constitute an update, with a dataset a hundred times larger, of the initial measurements~\cite{wzpaper} 
performed by ATLAS~\cite{atlaspaper}. 
The uncertainty on the integrated luminosity is reduced from $11\%$ to $3.4\%$ and the results
have significantly lower experimental systematic uncertainties. In addition, the $\Zee$ cross 
section is extended in lepton pseudorapidity ($\eta$) from a maximum of $2.47$ to $4.9$, 
increasing the range of parton momentum fraction $x$ probed. Measurements of the $W$ and 
$Z$ cross sections are presented in the electron and muon decay modes. The combination of 
the electron and muon inclusive cross section results and the $W/Z$ cross section ratio are compared 
to available predictions based on NNLO QCD using various PDF sets. In addition, the observation 
of $W$ and $Z$ bosons in the tau decay channel~\cite{wtaunu,ztautau_dilep,ztautau_lephad} 
provides a consistency check in the third lepton channel and constitutes a first characterization 
of an important background to new physics searches. 

\section{W and Z Cross Sections and Ratios in the Electron and Muon Channels}

The data were acquired during the LHC runs at $\sqrt{s} = 7$ TeV in $2010$, amounting
to an integrated luminosity of about $35~\ipb$, after requiring a fully operational detector in
stable beam conditions.
Events were selected for recording and subsequent offline analysis using a three-level single 
electron or muon trigger. 
A primary vertex (PV) with at least three tracks serves to select collision candidates. 
Electron identification is based on a sliding window algorithm forming calorimeter clusters
that are then associated to tracks in the Inner Detector (ID). 
To select  $\Wen$ events, exactly one electron cluster is required to have transverse 
energy $\et > 20$ GeV and to lie within the region where the ID acceptance is matched
to the precision measurements of the calorimeter: $|\eta|<2.47$, excluding the 
transition region between the barrel and endcap calorimeters $1.37<|\eta|<1.52$. 
The electron is required to satisfy different types of identification (``tight'') criteria to 
efficiently reject backgrounds from jets.
The missing transverse energy ($\met$) is based on the sum of calorimeter cell energy 
deposits in topological clusters and is required to be larger than $25$ GeV. 
Further, the transverse mass ($\mt$) is required to be larger than $40$ GeV. 
For $\Zee$ candidate events, two electrons are required to satisfy the less stringent 
``medium'' identification criteria, have opposite charge and an invariant mass in the range 
$66<m_{\mathrm{ee}}<116$ GeV. In the case of $\Zee$ events with an 
extended range of pseudorapidity, a central electron ($|\eta|<2.47$) passing the ``tight'' criteria is required and 
another electron has to be reconstructed in the forward region ($2.5 < |\eta| < 4.9$). The 
forward region extends beyond the ID coverage, so the electron identification is based on 
calorimeter cluster shapes alone. 

Muon identification is based on combined tracking using both the ID and the muon
spectrometer. Combined tracks with transverse momentum $\pt > 20$ GeV are selected
within $|\eta|<2.4$. Cosmic ray contamination is reduced with pointing requirements to the PV
in the center of the detector. Muons from heavy quark decays are removed with an isolation 
requirement relative to the muon $\pT$ of $\Sigma \pt^{\mathrm{ID}}/ \pT < 0.2$, where the ID 
tracks are in a cone in $\eta-\phi$ space of radius $0.4$ around the muon.
$\Wmn$ candidate events are selected requiring the $\MET$, determined from 
calorimeter and muon tracking information, to be larger than $25$ GeV and $\mT > 40$ GeV.
For $\Zmm$ candidate events, two muons are required to be of opposite charge and have 
an invariant mass in the range $66<m_{\mu \mu}<116$ GeV. 

The candidate sample size for each lepton channel is about $130,000$ for $W$ and $10,000$ 
for $Z$, with an additional $4,000$ forward $\Zee$ events. 
The electroweak background contributions to $W$ and $Z$ are obtained from Monte
Carlo simulation scaled to the NNLO cross section. The main background contributions to 
$W$ arise from $\Wtn$ ($2.8\%$), where the $\tau$ decays leptonically, and 
additionally in the muon channel from $\Zmm$ events, where one of the muons falls outside 
of the acceptance ($3.5\%$). The electroweak background contributions to $\Zee$ and $\Zmm$ 
are small ($0.3\%$ and $0.1\%$ respectively), as are the diboson contributions ($0.2\%$).
The background from jets produced via QCD processes (referred to as ``QCD background'') is 
derived from fits comparing the $\MET$ (for $W$) or the invariant mass (for $Z$) distributions 
in data and in templates derived from simulation. The resulting predictions are of $2-4\%$ for 
the electron channel analyses, except for the case of the forward $\Zee$ selection ($27\%$), 
where $W$s in association with jets faking a forward electron provide an additional background. 
The QCD background in the muon channel is estimated from extrapolations from control regions, 
yielding contributions of $1.7\%$ for $\Wmn$ and $0.2\%$ for $\Zmm$.

\begin{figure}[!h]
  \includegraphics[width=.5\textwidth]{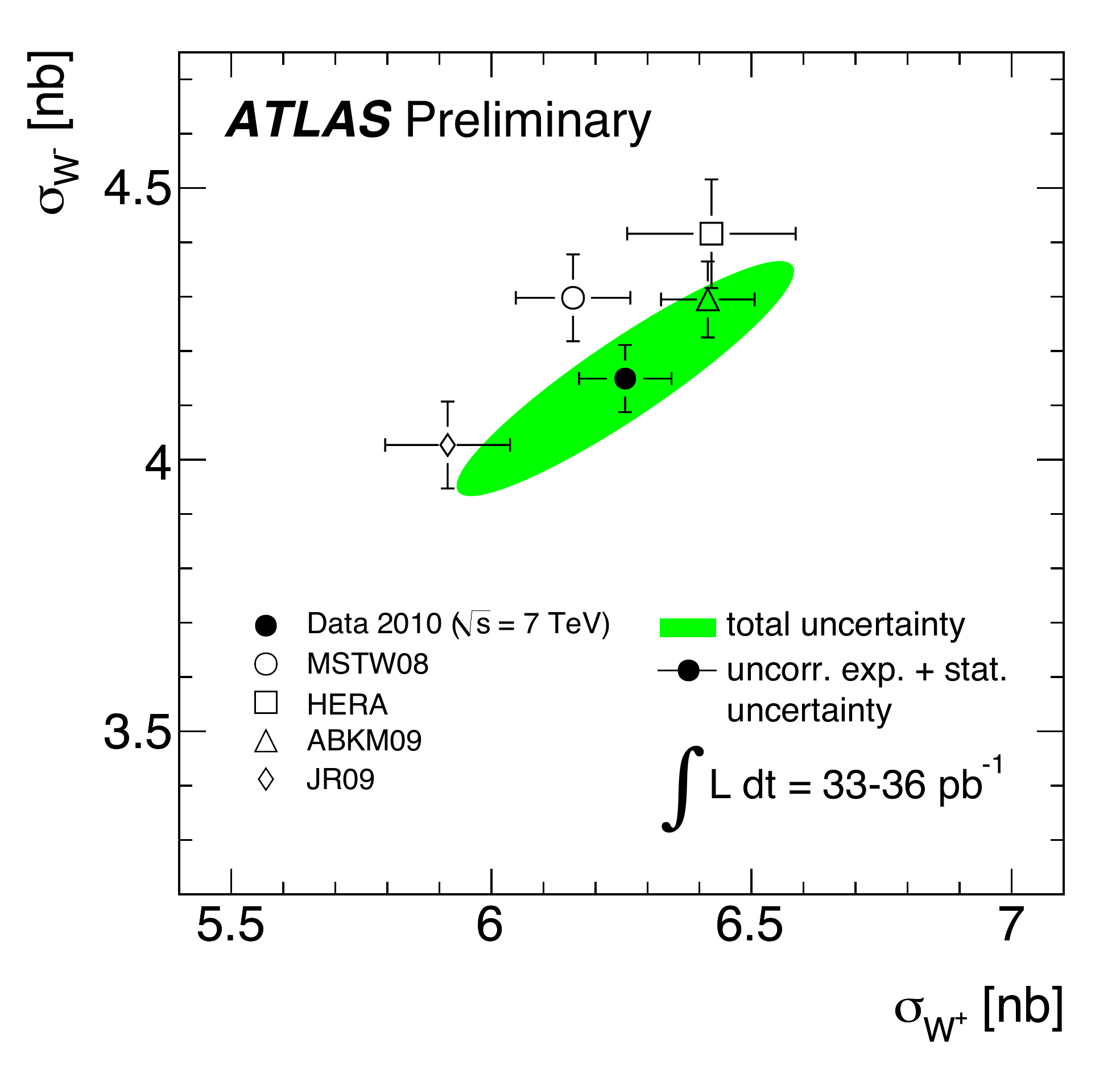}%
  \includegraphics[width=.5\textwidth]{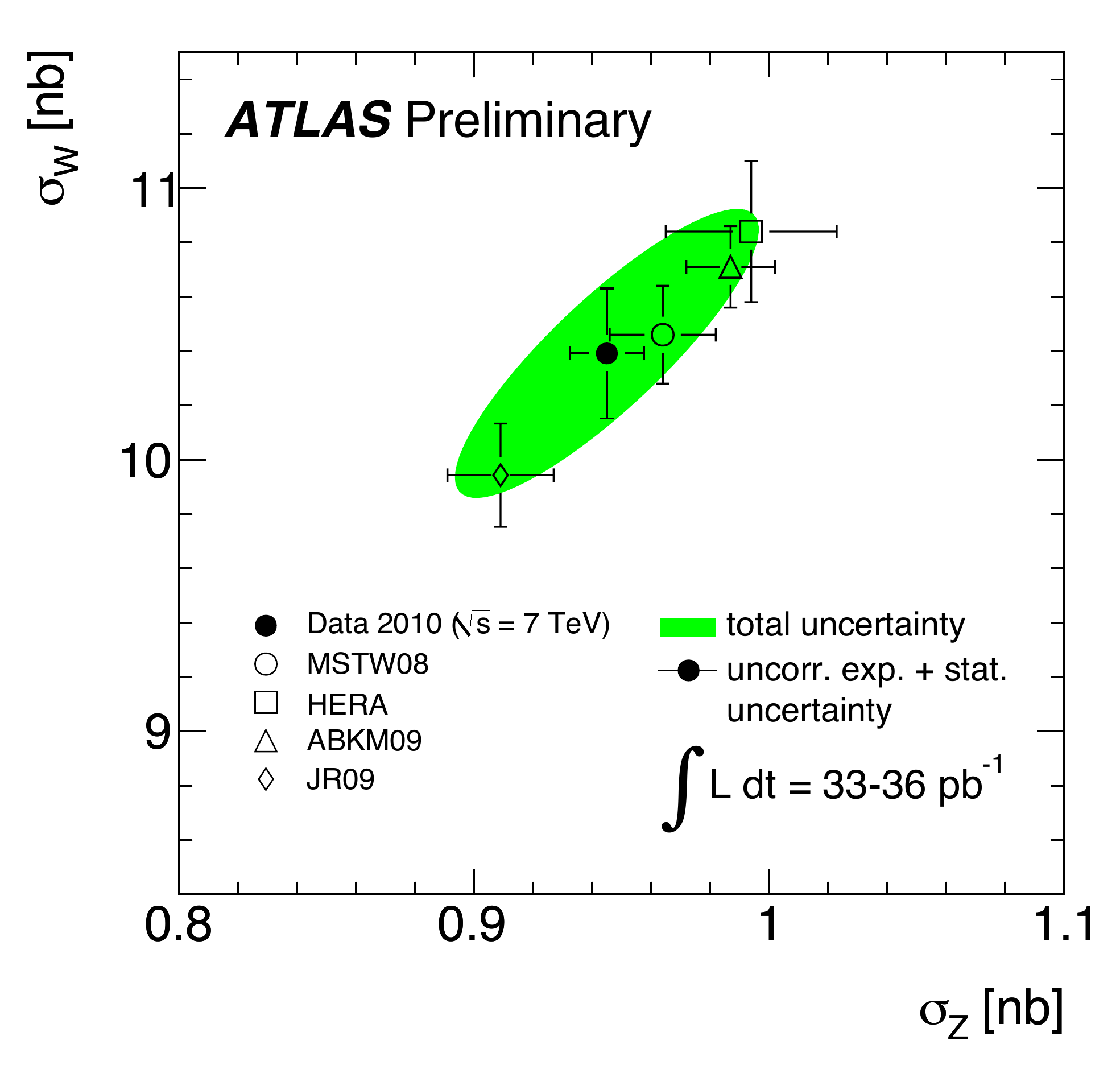}
 \caption{Measured and predicted cross sections times leptonic branching ratios, 
    for $\sigma_{W^-}$ vs. $\sigma_{W^+}$ (left) and $(\sigma_{W^+} + \sigma_{W^-})$ 
    vs. $\sigma_{Z / \gamma^*}$ (right). 
   The correlations between the various cross sections are accounted for and the projections 
   of the ellipse to the axes correspond to one standard deviation uncertainty of the cross 
   sections. Theoretical predictions are calculated to NNLO in QCD using PDFs by the 
   MSTW~\cite{mstw}, ABKM~\cite{abkm09, abkm10}, HERA~\cite{hera, herannlo} and 
   JR~\cite{jr} groups. Only the PDF uncertainties on the theoretical predictions are shown.}
 \label{fig:wzxsec}
\end{figure}

The total $W$ and $Z$ cross section times branching ratio is measured
using a ``cut-and-count'' method, whereby candidate events are selected in data, the number
of expected background events is subtracted and the signal yield is divided by the 
integrated luminosity corresponding to the run selections and trigger employed. This result is 
corrected for acceptances and detector effects. Measurements of the integrated cross sections 
within the fiducial regions, providing the most precise experimental results for comparison to 
theory, and after extrapolation to the full kinematic region can be found in~\cite{wzconf} for 
the electron and muon channels. 
The total inclusive cross sections are consistent for the two decay modes within
their measurement uncertainties. The combination of both channels is shown in 
Table~\ref{tab:xseccomb} and Figure~\ref{fig:wzxsec}, assuming the two lepton decay 
mode cross sections to be independent, with the exception of the common parts in the 
acceptance calculation, the $\MET$ uncertainty in the $W$ channels and the luminosity 
uncertainty that are treated as fully correlated. 

\begin{table}[!h]
\begin{tabular}{lrc}
\hline
\tablehead{1}{c}{b}{Process} & \tablehead{1}{c}{b}{Combined Measurement} & \tablehead{1}{c}{b}{MSTW08}\\
\hline
$\mathbf{\sigma^{\mathrm{tot}}_{W+} \cdot \mathrm{BR(\Wln)}}$ [nb]
& $6.257\pm0.017\pm0.152\pm0.213\pm0.188$ & $6.16\pm 0.11$ \\
$\mathbf{\sigma^{\mathrm{tot}}_{W-} \cdot \mathrm{BR(\Wln)}}$ [nb]
& $4.149\pm0.014 \pm 0.102 \pm 0.141 \pm 0.124$& $4.30 \pm 0.08$ \\
$\mathbf{\sigma^{\mathrm{tot}}_{W} \cdot \mathrm{BR(\Wln)}}$ [nb]
& $10.391\pm0.022 \pm 0.238 \pm 0.353 \pm 0.312$& $10.46 \pm 0.18$ \\
$\mathbf{\sigma^{\mathrm{tot}}_{Z/ \gamma*} \cdot \mathrm{BR(Z/ \gamma* \rightarrow ll)}}$ [nb]
& $0.945 \pm 0.006 \pm 0.011 \pm 0.032 \pm 0.038$& $0.964 \pm 0.018$ \\
\hline
$\sigma^{\mathrm{tot}}_{W+} \cdot \mathrm{BR}/\sigma^{\mathrm{tot}}_{Z} \cdot \mathrm{BR}$ 
& $6.563\pm0.049\pm0.134\pm0.098$& $6.388\pm0.026$ \\
$\sigma^{\mathrm{tot}}_{W-} \cdot \mathrm{BR}/\sigma^{\mathrm{tot}}_{Z} \cdot \mathrm{BR}$ 
& $4.345\pm0.034\pm0.095\pm0.065$& $4.459\pm0.018$ \\
$\sigma^{\mathrm{tot}}_{W} \cdot \mathrm{BR}/\sigma^{\mathrm{tot}}_{Z} \cdot \mathrm{BR}$ 
& $10.906\pm0.079\pm0.215\pm0.164$& $10.85\pm0.018$ \\
\hline
\end{tabular}
\caption{Total averaged cross sections times leptonic branching ratios in nb
 for $W^+$, $W^-$, $W$ and $Z/ \gamma^*$ production in the
 combined electron and muon final states. The uncertainties
 denote the statistical, experimental systematic, luminosity and acceptance errors. 
 The ratios of the cross sections times leptonic branching fractions are also shown with
 their corresponding statistical, experimental systematic and acceptance uncertainties. 
 Theoretical predictions for one choice of PDF set~\cite{mstw} at NNLO in perturbative QCD are included 
 for comparison, showing the PDF uncertainties only, taken at $68\%$ C.L.}
\label{tab:xseccomb}
\end{table}

Compared to the first measurement in ATLAS~\cite{wzpaper}, the statistical uncertainty 
of the total cross-section measurement is improved to the level of $<1\%$, the systematic 
uncertainty by a factor of about three, to $2.4\%$ and $1.2\%$ for the $W$ and $Z$ cross 
sections respectively, and the luminosity uncertainty by a factor of four to $3.4\%$. The 
$\Wen$ cross sections are measured with an experimental uncertainty of $2.8\%$ to $3.0\%$ 
for the two charges, which contains approximately equal contributions from the combined 
electron efficiency and the missing transverse energy. The central $\Zee$ cross section is 
measured with an experimental precision of $3.5\%$, dominated by the uncertainty of the 
reconstruction efficiency measurement. The $\Zee$ using the forward selection has an 
experimental precision of about $10\%$. Within their uncertainties the central and forward 
$Z$ results are compatible. 
The $\Wmn$ cross section is measured with an experimental uncertainty of $2.4\%$ and 
of $2.7\%$ for $\Wplus$ and $\Wminus$ respectively, also dominated by the 
contribution from the calorimeter $\MET$ scale and resolution ($2\%$). 
The $\Zmm$ cross section is measured with an experimental precision of $1.1\%$, mainly 
arising from the reconstruction efficiency uncertainty. 
The uncertainty on the theoretical extrapolation from the fiducial region of the measurement 
to the total phase space is $3.0\%$ and $4.0\%$ in the $W$ and $Z$ measurements 
respectively~\cite{wzpaper}.
The cross section results in Figure~\ref{fig:wzxsec} are compared with predictions from 
four sets of NNLO parton distributions, showing consistency of all NNLO predictions within 
the total accuracy of the measurement of about $5\%$. The ratios of the measured $W$ and 
$Z$ cross sections are also listed in Table~\ref{tab:xseccomb}, accounting for the correlations 
of errors. The uncertainty on these ratios ($2.7\%$  for $W/Z$) is smaller than that on the 
cross sections due to cancellations in the ratios and the results are also consistent with NNLO 
predictions.

\section{W and Z Observation in the Tau Channel}

The first observation of $\Wtn$ events with a hadronic $\tau$ candidate and large $\MET$
has been performed with a dataset corresponding to $546~\inb$ of integrated luminosity~\cite{wtaunu}. 
In addition, $\Ztt$ events have been observed in three decay channels: the dilepton
decay mode $\Ztt \rightarrow e\mu+4\nu$ using a dataset of $35~\ipb$~\cite{ztautau_dilep} 
and in the lepton-hadron decay modes including a soft electron or muon, using $8.3~\ipb$ 
and $8.5~\ipb$ of integrated luminosity, respectively~\cite{ztautau_lephad}. The signal yields 
are in agreement with Standard Model expectations for all $W$ and $Z$ analyses.




\bibliographystyle{aipproc}   

\bibliography{DISProc_MartinezOutschoorn}




{\it \small ~~~~~~~~~~~~~~~~~~~~~~Copyright CERN for the benefit of the ATLAS collaboration}

\end{document}